\documentclass{article}
\usepackage{spconf,amsmath,epsfig}
\usepackage{graphicx,amssymb,booktabs, subfigure}
\usepackage{verbatim}
\let\OLDthebibliography\thebibliography
\renewcommand\thebibliography[1]{
  \OLDthebibliography{#1}
  \setlength{\parskip}{0pt}
  \setlength{\itemsep}{0pt plus 0.3ex}
}

\begin{document}\sloppy

\def\x{{\mathbf x}}
\def\L{{\cal L}}

\title{MULTI-SCALE TEMPORAL-FREQUENCY ATTENTION FOR MUSIC SOURCE SEPARATION}
%
\name{Lianwu Chen, Xiguang Zheng, Chen Zhang, Liang Guo, Bing Yu}
\vspace{-10pt}
\address{Kuaishou Technology Co. Beijing, China \\
chenlianwu@kuaishou.com}
\maketitle
\vspace{-15pt}

\begin{abstract}

In recent years, deep neural networks (DNNs) based approaches have achieved the start-of-the-art performance for music source separation (MSS). Although previous methods have addressed the large receptive field modeling using various methods, the temporal and frequency correlations of the music spectrogram with repeated patterns have not been explicitly explored for the MSS task. In this paper, a temporal-frequency attention module is proposed to model the spectrogram correlations along both temporal and frequency dimensions. Moreover, a multi-scale attention is proposed to effectively capture the correlations for music signal. The experimental results on MUSDB18 dataset show that the proposed method outperforms the existing state-of-the-art systems with 9.51 dB signal-to-distortion ratio (SDR) on separating the vocal stems, which is the primary practical application of MSS.

\end{abstract}
\section{Introduction}
\label{sec:intro}
During music production, recordings of vocals and individual instruments called stems are mixed together into the final song. Music source separation (MSS) is designed to separate the mixed signal into the individual stems. Since the separated stems can be used in various application such as Karaoke systems \cite{rafii2012repeating} or music up-mixing \cite{fitzgerald2011upmixing}, MSS has received increasing interest in recent years. As a subtask of the Signal Separation Evaluation Campaign (SiSEC), the separated stems of MSS were categorized into vocals, bass, drums and other \cite{stoter20182018}. 

While traditional approaches have been proposed in \cite{lee1999learning,davies2007source},  methods based on deep neural networks (DNNs) have outperformed these traditional approaches in recent years.
In \cite{uhlich2015deep, nugraha2016multichannel}, neural network with several fully connected layers was utilized to separate the audio sources. To capture the temporal context, features of multiple frames were concatenated as the input of network. In \cite{uhlich2017improvingLSTM,stoter2019openLSTM,liu2019dilatedGRU}, recurrent neural networks were used to capture the longer temporal contexts. In most of recent works \cite{takahashi2018mmdenselstm,takahashi2020d3net,defossez2019musicDemus,hennequin2020spleeter,kong2021decoupling}, Convolutional Neural Network (CNN) based encoder-decoder architecture were employed and have achieved the state-of-the-art performance. By stacking several 2-dimensional CNN layers, the model can capture both temporal and frequency context. To obtain large temporal-frequency receptive field with high efficiency, one of the popular operations is repeatedly resampling the feature maps \cite{jansson2017singingSpecUnet,stoller2018wave,takahashi2017multi}. More specifically, the feature maps are downsampled repeatedly in the encoder, so that the CNN layers in lower resolution representation can obtain larger receptive field. 
Then these low resolution feature maps are upsampled repeatedly in the decoder to obtain the same resolution of input feature. 

Besides, several additional modules are proposed to further capture the temporal and frequency context for the encoder-decoder architecture. In \cite{takahashi2018mmdenselstm, defossez2019musicDemus}, LSTM layers were added between the encoder and the decoder to efficiently model long-term musical structures. In \cite{choi2019investigating}, a time-distributed fully-connected network was proposed to extract the long-range correlations existed along the frequency axis. In \cite{takahashi2020d3net}, multi-dilated convolution with different dilation factors is utilized to model different resolution and obtain larger temporal and frequency receptive field. In \cite{kong2021decoupling}, sufficiently large receptive field was obtained by a residual UNet architecture with up to 143 layers, and the system achieves the state-of-the-art MSS performance.

\begin{figure*}[htb]
  \centering
  \includegraphics[width=1.0\linewidth,height=6.5cm]{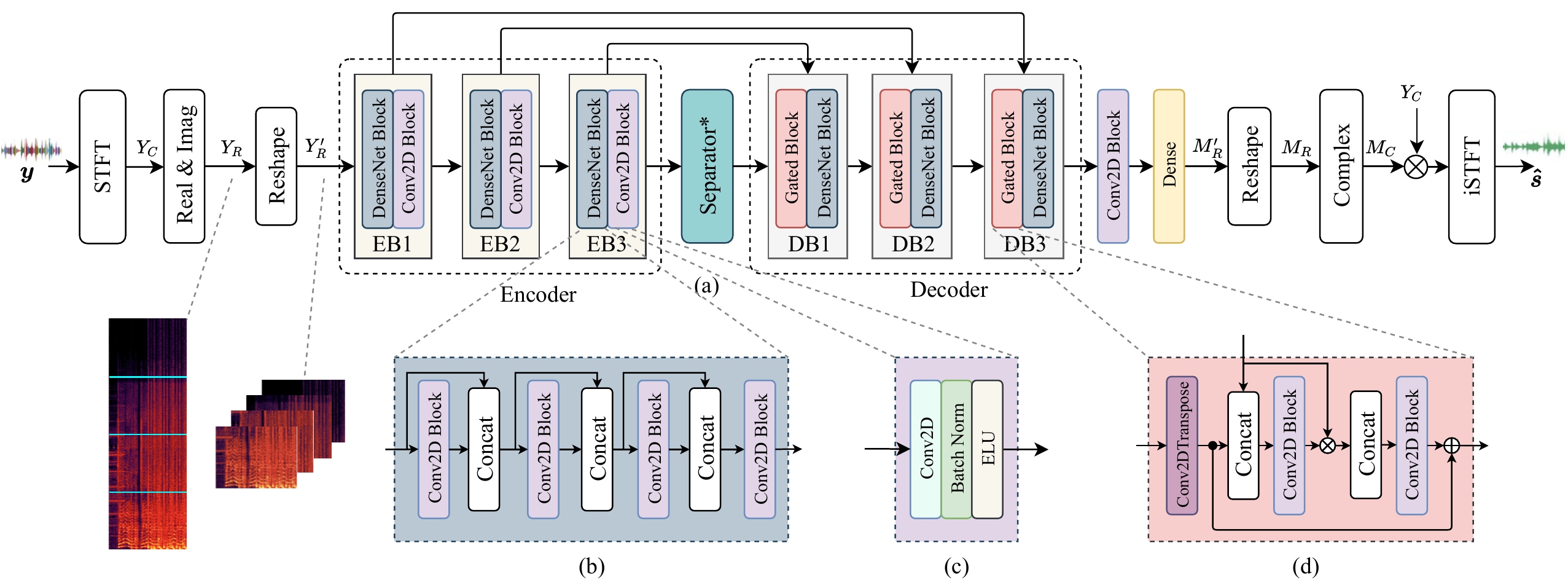}\\
  \vspace{-6pt}
  \caption{(a) The proposed system architecture.(Separator* is shown in Fig.2 and Fig.3) (b) The structure of a DenseNet Block. (c) The structure of a Conv2D Block. (d) The structure of a Gated Block in the decoder.}
  \label{fig1}
  \vspace{-6pt}
\end{figure*}

 While most of the existing MSS systems can model large receptive field, the correlation along temporal or frequency dimension has not been explicitly exploited. This is especially crucial for the MSS task \cite{paulus2010state}, for example, the temporal correlation in beat and downbeat patterns, and the frequency correlation in chorus, harmony and chords. In \cite{liu2020voiceAccoSepSelfAtt, li2021samsSlicedAttMSS} self-attention was used to exploit long-term dependency of music signal, but they only considered the attention along the temporal dimension. Motivated by the success of temporal and frequency self-attention mechanism in speech enhancement task \cite{SNNET}, a new separation module with temporal and frequency self-attention layers is proposed to capture the spectrogram correlations within the encoder-decoder based MSS architecture. Moreover, considering the different  frequency ranges of various instruments and rapid changes of music content, a multi-scale mechanism is introduced to capture the correlations, improving the robustness of proposed method on various music styles. 
Compared to the mainstream MSS systems, the proposed method  also provides a new way to obtain large receptive field without may repeated resample layers or other additional modules.

The contributions of this paper can be summarized as follows: 1). We propose a temporal-frequency attention layers in encoder-decoder based architecture to capture the spectrum correlations for MSS. 2). We further introduce a multi-scale mechanism to effectively model spectrum correlations for different temporal and frequency ranges. 3). We experimentally show the effectiveness of the proposed systems, which achieves start-of-the-art results on MUSDB18 dataset \cite{stoter20182018}.

\section{PROPOSED SYSTEM}
\vspace{-6pt}
\subsection{Overview}
As shown in Fig.\ref{fig1}(a), the neural network consists of an encoder, a separator and a decoder. It takes a discrete stereo signal with $N$ samples for each channel $\boldsymbol{y}\in \mathbb{R}^{N \times 2}$ as the input. The input signal is then transformed into a time-frequency domain representation $Y_C \in \mathbb{C}^{T \times F \times 2}$ via STFT, where $T$ is the number of frames and $F$ is the number of frequency bins of the complex spectrogram; 2 refers to the two-channel stereo input. To form the input of the neural network, the real and imaginary components are concatenated as $Y_R=[real(Y_C); imag(Y_C)] \in \mathbb{R}^{T \times F \times 4}$.

To reduce the computational cost for high-resolution input, following the subband mechanism proposed in \cite{liu2020channel}, the fullband spectrogram is sliced into $K$ subbands to form the channel-wise subband signal $Y'_R \in \mathbb{R}^{T \times (F/K) \times (4 \times K)}$ served as the encoder input, where $K$ is set to 4 according to our preliminary results. 
The neural network estimates a channel-wise subband mask $M'_R\in \mathbb{R}^{T \times (F/K) \times (4 \times K)}$ and then reshaped to the fullband mask $M_C\in \mathbb{C}^{T \times F \times 2}$ for the stereo complex spectrogram. Finally, the target signal is estimated by multiplying the mixture signal $Y_C$ with the estimated mask $M_C$. The time-domain target signal $\hat{\boldsymbol{s}}$ can be obtained via iSTFT.

\vspace{-6pt}
\subsection{Encoder and Decoder}
\label{ssec:encdec}
As shown in Fig.\ref{fig1}(a), the encoder has three encoder blocks (EBs), each consists of a DenseNet block (detailed in  Fig.1(b)) and a Conv2D block (detailed in  Fig.1(c)). The DenseNet block consists of four Conv2D blocks (detailed in  Fig.1(c)) with concatenation operations. The DenseNet block can learn explicit cross-layer interactions and reuses features computed in preceding layers, which yields efficient parameter utilization and suits for the MSS problems as discussed in \cite{takahashi2017multi}. The Conv2D block consists of a Conv2D layer, a Batch Normalization layer and an ELU activation layer.

The decoder consists of three decoder blocks (DBs) followed by a Conv2D block and a Dense layer. Each decoder block consists of a gated block \cite{SNNET} and a DenseNet block. The gated block consists of a Conv2DTanspose layer and two Conv2D blocks (detailed in Fig.\ref{fig1}(c)) as shown in Fig.\ref{fig1}(d), which learns a multiplicative mask for the feature from the encoder and suppress its undesired part. The structure of the Dense block in decoder is identical to the one used in encoder. After a Conv2D block, a Dense layer with tanh activation is utilized to generate the real and imaginary components of the complex ideal ratio mask (cIRM) with boundary of [-1, 1]. 
The cIRM is further extended to [-2, 2] by multiplying a expanding factor of 2 to increase the upper bound of the oracle SDR as discussed in \cite{kong2021decoupling}. The expanding factor is chosen to maximize the separation SDR according to our preliminary experiments. The detailed hyper-parameters of the encoder and decoder is listed in Table \ref{Conv2D_layers_configuration}. 

\begin{figure*}[ht]
  \centering
  \includegraphics[width=1.0\linewidth]{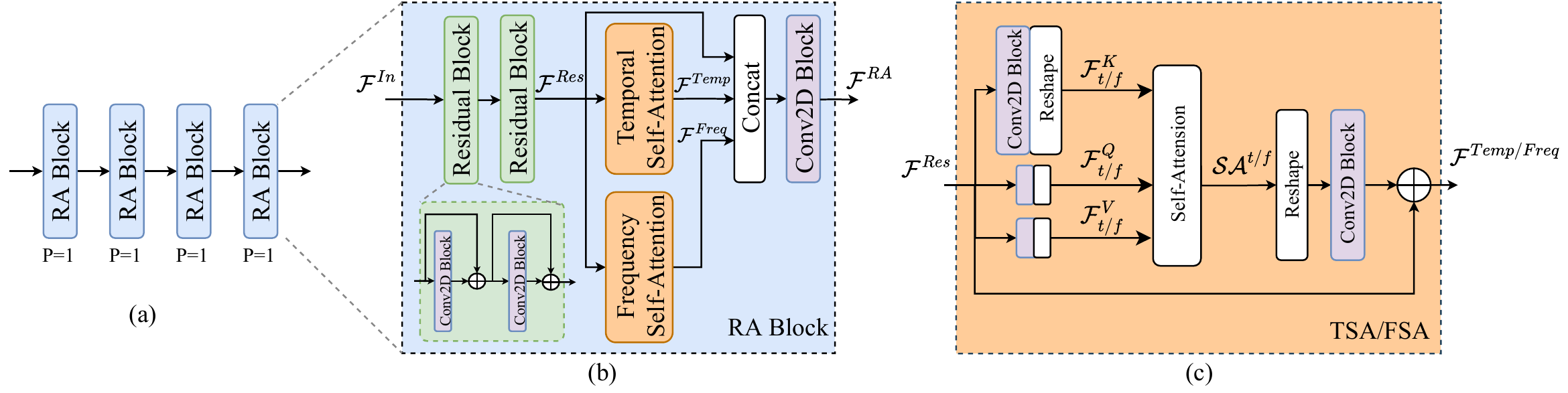}\\
  \vspace{-8pt}
  \caption{(a) Separator with temporal-frequency attention. (b) The structure of the residual attention (RA) block. (c) The structure of  temporal self-attention and frequency self-attention.}
  \label{fig.Separator}
\end{figure*}
\begin{table}[htb]
  \caption{The configurations of encoder and decoder}
  \label{Conv2D_layers_configuration}
  \centering
  \small
  \begin{tabular}{l l | c c c}
    \toprule
    & \textbf{Layer} & \textbf{Channel} & \textbf{Kernel} & \textbf{Stride} \\
    \midrule
    $\text{EB1}$ &  $\text{DenseNet}$  & $32$ & $(3,3)$ & $(1,1)$~~~ \\
       &  $\text{Conv2D}$  & $32$ & $(3,3)$ & $(1,1)$~~~ \\
    \midrule
    $\text{EB2}$ &  $\text{DenseNet}$  & $64$ & $(3,3)$ & $(1,1)$~~~ \\
       &  $\text{Conv2D}$  & $64$ & $(3,3)$ & $(2,2)$~~~ \\
    \midrule
    $\text{EB3}$ &  $\text{DenseNet}$  & $64$ & $(3,3)$ & $(1,1)$~~~ \\
       &  $\text{Conv2D}$  & $64$ & $(3,3)$ & $(1,2)$~~~ \\
    \midrule
    $\text{DB1}$ &  $\text{Conv2DTranspose}$  & $64$ & $(3,3)$ & $(1,2)$~~~ \\
    &  $\text{Conv2D}$  & $64$ & $(1,1)$ & $(1,1)$~~~ \\
    &  $\text{DenseNet}$  & $64$ & $(3,3)$ & $(1,1)$~~~ \\
    \midrule
    $\text{DB2}$ &  $\text{Conv2DTranspose}$  & $64$ & $(3,3)$ & $(2,2)$~~~ \\
    &  $\text{Conv2D}$  & $64$ & $(1,1)$ & $(1,1)$~~~ \\
    &  $\text{DenseNet}$  & $64$ & $(3,3)$ & $(1,1)$~~~ \\
    \midrule
    $\text{DB3}$ &  $\text{Conv2DTranspose}$  & $32$ & $(3,3)$ & $(1,1)$~~~ \\
    &  $\text{Conv2D}$  & $32$ & $(1,1)$ & $(1,1)$~~~ \\
    &  $\text{DenseNet}$  & $32$ & $(3,3)$ & $(1,1)$~~~ \\
    \midrule
    &  $\text{Conv2D}$  & $4 \times K$ & $(1,1)$ & $(1,1)$~~~ \\
    \bottomrule
  \end{tabular}
  \vspace{0pt}
\end{table}

\subsection{Temporal-Frequency Attention based Separator}
\label{ssec:att}

As shown in Fig.\ref{fig.Separator}(a), the temporal-frequency attention based separator consists of four residual attention (RA) blocks (detailed in Fig.\ref{fig.Separator}(b)). The input feature map of the RA block is $\mathcal{F}^{In}\in \mathbb{R}^{T' \times F' \times C}$ generated from encoder or previous RA block. $T'$ is the time steps. $F'$ is the feature dimension. $C$ is equal to 64. The input feature map is fed into two residual blocks. Each residual block consists of two Conv2D blocks with a kernel size of (3,3) and a stride of (1,1). The output of the residual block $\mathcal{F}^{Res}\in \mathbb{R}^{T' \times F' \times C}$ is then fed parallel into the temporal self-attention (TSA) and the frequency self-attention (FSA) blocks to capture the global dependencies along temporal and frequency dimensions, respectively. The outputs of the two self-attention blocks $\mathcal{F}^{Temp}\in \mathbb{R}^{T' \times F' \times C}$ and $\mathcal{F}^{Freq}\in \mathbb{R}^{T' \times F' \times C}$ are concatenated with $\mathcal{F}^{Res}$ to fed into a Conv2D block to generate the output of the RA block $\mathcal{F}^{RA}\in \mathbb{R}^{T' \times F' \times C}$, the kernel size and stride of the output Conv2D are all (1,1).       

The TSA and FSA blocks share the same structure with different reshape operations as shown in Fig.\ref{fig.Separator}(c). The input $\mathcal{F}^{Res}\in \mathbb{R}^{T' \times F' \times C}$ is fed parallel into Conv2D blocks. The kernel size and stride of the Conv2D blocks are all (1,1), the channel number is reduced by half to $\frac{C}{2}$ for less computational complexity.
The output feature maps of the Conv2D blocks in $\mathbb{R}^{T' \times F' \times \frac{C}{2}}$ are then reshaped to the $\mathcal{F}_{t}^{k} \in \mathbb{R}^{T' \times (\frac{C}{2} \times F')}$ for TSA or $\mathcal{F}_{f}^{k} \in \mathbb{R}^{F' \times (\frac{C}{2} \times T')}$ for FSA, respectively, where $k\in [{K,Q,V}]$. $K,Q,V$ indicates the key, query and value in the scaled dot-product self-attention \cite{attention}.
For TSA, the self-attention is formulated as:
\begin{equation}
SA^{t}=Softmax(\mathcal{F}_{t}^{Q} \cdot (\mathcal{F}_{t}^{K})^{H}/\sqrt{\frac{C}{2} \times F^{'}})\cdot\mathcal{F}_{t}^{V}
\label{t_attention_2}
\end{equation}
where $SA^{t}\in \mathbb{R}^{T' \times (\frac{C}{2} \times F')}$, $()^{H}$ denotes matrix transpose and $\cdot$ denotes matrix multiplication. For FSA, the self-attention is formulated as:
\begin{equation}
SA^{f}=Softmax(\mathcal{F}_{f}^{Q} \cdot (\mathcal{F}_{f}^{K})^{H}/\sqrt{\frac{C}{2} \times T'})\cdot\mathcal{F}_{f}^{V}
\label{f_attention_2}
\end{equation}
where $SA^{f}\in \mathbb{R}^{F' \times (\frac{C}{2} \times T')}$. 
The $SA^{t}$ and $SA^{f}$ are further reshaped to $\mathbb{R}^{T' \times F' \times \frac{C}{2}}$ and then fed into a Conv2D block with channel number $C$, kernel size (1,1) and stride (1,1). The input $\mathcal{F}^{Res}$ is added to the output of the Conv2D block to get the final temporal self-attention $\mathcal{F}^{Temp}$ or frequency self-attention $\mathcal{F}^{Freq}$.

\subsection{Multi-scale Temporal-Frequency Attention}
In section \ref{ssec:att}, the temporal attention is calculated based on all frequency bins. The frequency attention is calculated based on all input time steps. However, considering the different frequency ranges of various instruments and rapid changes of music content, using all frequency or temporal features might not be the optimal way for attention calculation. Therefore, multi-scale  segment-wise attention is proposed to calculate attention based on different frequency or temporal ranges. More specifically, the input of TSA and FSA is first sliced into $P$ segments along the frequency and temporal dimension respectively. The attentions are then calculated for each segment individually and combined into the final output. The segment-wise TSA is formulated as:
\begin{figure}[htb]
  \centering
  \includegraphics[width=0.6\linewidth,height=3.5cm]{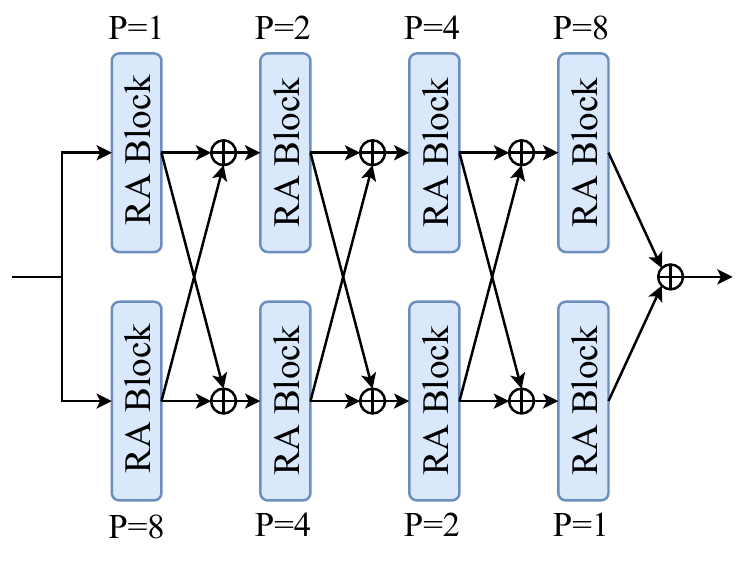}\\
  \vspace{-10pt}
  \caption{Separator with  multi-scale temporal-frequency attention.}
  \label{fig.SeparatorMS}
  \vspace{-8pt}
\end{figure}
\begin{equation}
\mathcal{F}^{Temp}=Concat(\mathcal{F}_{FP}^{Temp}(1), ..., \mathcal{F}_{FP}^{Temp}(P))
\label{t_attention_MS1}
\end{equation}
\begin{equation}
\mathcal{F}_{FP}^{Temp}(i)=TSA(\mathcal{F}_{FP}^{Res}(i)))
\label{t_attention_MS2}
\end{equation}
where the $\{\mathcal{F}_{FP}^{Temp}(i),\mathcal{F}_{FP}^{Res}(i)\} \in \mathbb{R}^{T' \times \frac{F'}{P} \times C}$ is the i-th segment of $\mathcal{F}_{FP}^{Temp}$ and $\mathcal{F}_{FP}^{Res}$ respectively. The subscript $FP$ indicates that the $\mathcal{F}^{Temp}$ or $\mathcal{F}^{Res}$ is sliced into $P$ segments along frequency dimension. $TSA()$ is the temporal self-attention module as shown in Fig.\ref{fig.Separator}(c). The segment-wise FSA is formulated as:
\begin{equation}
\mathcal{F}^{Freq}=Concat(\mathcal{F}_{TP}^{Freq}(1), ..., \mathcal{F}_{TP}^{Freq}(P))
\label{f_attention_MS1}
\end{equation}
\begin{equation}
\mathcal{F}_{TP}^{Freq}(i)=FSA(\mathcal{F}_{TP}^{Res}(i)))
\label{f_attention_MS2}
\end{equation}
where the $\{\mathcal{F}_{TP}^{Freq}(i),\mathcal{F}_{TP}^{Res}(i)\} \in \mathbb{R}^{\frac{T'}{P}  \times F' \times C}$  is the i-th segment of $\mathcal{F}_{TP}^{Freq}$ and $\mathcal{F}_{TP}^{Res}$ respectively. The subscript $TP$ indicates that the $\mathcal{F}^{Freq}$ or $\mathcal{F}^{Res}$ is sliced into $P$ segments along temporal dimension.

The RA blocks with different value of $P$ can be combined together to form a multi-scale mechanism. Fig.\ref{fig.Separator}(a) shows the single-scale attention consists of four RA blocks with $P=1$. Fig.\ref{fig.SeparatorMS} shows the separator with  multi-scale attention, which consists of RA blocks with $P=1,2,4,8$. To obtain both small and large scale attention at the same layer, a two-branch structure with parallel RA blocks is introduced, where one branch contains RA blocks with increasing $P$ value while the other contains RA blocks with decreasing $P$ value.

\subsection{Loss Function}
A joint loss function with combining time domain and frequency domain losses is employed to train the network. The time domain loss is defined as mean absolute error (MAE) between target signal $\boldsymbol{s}$ and estimated signal $\hat{\boldsymbol{s}}$,
\begin{equation}
  \mathcal{L}_{time} = ||\boldsymbol{s} - \hat{\boldsymbol{s}}||_1
\end{equation}
Where $\left|| \cdot \right||_1$ denotes L1-Norm. The frequency domain loss is defined as MAE between target complex spectrum $\boldsymbol{S}$ and estimated complex spectrum $\hat{\boldsymbol{S}}$, 
\begin{equation}
  \mathcal{L}_{freq} = ||real(\boldsymbol{S} - \hat{\boldsymbol{S}})||_1 + ||imag(\boldsymbol{S} - \hat{\boldsymbol{S}})||_1
\end{equation}
The overall loss is defined as,
\begin{equation}
  \mathcal{L} = \mathcal{L}_{time} + \alpha \cdot \mathcal{L}_{freq} 
\end{equation}
where $\alpha$ is set to 0.1 based on preliminary experiment results.

\section{Experiments}
\label{sec:Experiments}

\subsection{Dataset and setup}
\label{ssec:Dataset}
The proposed system is evaluated based on the MUSDB18 dataset  \cite{stoter20182018}, which consists of 150 songs with stereo format and 44.1kHz sampling rate. For each song, the final mixture signal is provided with its four audio stems, namely, vocals, bass, drums and other. We adopted the official split of 86, 14 and 50 songs for train, development, and evaluation respectively. 
The audio recordings are split into around 5.6 seconds (240 frames) segments with 2 seconds (86 frames) shift. The time domain segments are transformed to the time-frequency domain using an $8192$ samples STFT with $1024$ samples hop size. The complex spectrum $Y_C\in \mathbb{C}^{T \times F}$ of mixture clip is used as the input of system, where $T=240$ and $F=4096$. Data augmentation with channel swapping and remixing \cite{pretet2019singing} is employed on the fly during model training. 

For each audio source, we train a dedicated model individually. The Adam optimizer is employed. The initial learning rate is set to 0.001. It will decay by a factor of 0.8 when the validation loss does not decrease for 10 epochs. All the models are trained for 300 epochs.

\subsection{Comparison with the existing systems}
\label{ssec:ComparisonSOTA}
We compare the proposed Multi-scale Temporal-Frequency Attention Network (MTFAttNet) with other existing start-of-the-art systems on MUSDB18 dataset in Table \ref{SDR_STOA}. The signal-to-distortion ratio (SDR) \cite{vincent2006performance} computed by the museval toolbox\cite{stoter20182018} is used as evaluation metric.
In the upper half of Table \ref{SDR_STOA}, the SDR results of the existing single domain systems including  Spleeter\cite{hennequin2020spleeter}, D3Net\cite{takahashi2020d3net}, Demucs \cite{defossez2019musicDemus} and ResUNetDecouple+ \cite{kong2021decoupling} are compared with the proposed MTFAttNet system. The $W$ and $S$ indicates the waveform domain and spectrogram domain respectively. In the bottom half of Table \ref{SDR_STOA}, we also listed the results of the top-ranked hybrid domain systems in the Music  Demixing (MDX) challenge at ISMIR 2021 \cite{mdx2021}, namely, $\text{KUIELab-MDX-Net}$ \cite{kim2021kuielab} and  $\text{Hybrid Demucs}$\cite{defossez2021hybrid}. $W+S$ indicates the method is working on hybrid domain. 

As shown in Table \ref{SDR_STOA}, compared to the existing single domain methods, the proposed MTFAttNet method achieves significant improvement in separating vocals, drums and other stems. The bass SDR of the waveform domain method Demucs is higher than all the spectrogram domain methods, which might be caused by the limited frequency resolution of spectrum for bass. The overall SDR of the proposed system is 7.26 dB, which outperforms the best existing single domain method (ResUNetDecouple+ with SDR of 6.73 dB). Although the overall SDR is lower than the start-of-the-art hybrid domain systems, the proposed method still achieves the best vocal separation performance with the SDR of 9.51 dB, which outperforms the best existing hybrid domain method (KUIELab-MDX-Net with vocal SDR of 9.00 dB).

\begin{table}[htb]
  \caption{SDR comparison for the proposed and existing MSS systems.}
  \label{SDR_STOA}
  \centering
  \small
  \setlength\tabcolsep{1.5pt}
  \begin{tabular}{l c c c c c c}
    \toprule
   \textbf{Method} & \textbf{Domain} &\textbf{Vocals} & \textbf{Bass} & \textbf{Drums} & \textbf{Other} & \textbf{All} \\
    \midrule
    $\text{Spleeter}$ & $\text{S}$  & 6.86 & 5.51  &  6.71  & 4.55 &  5.91 \\
    $\text{D3Net}$ & $\text{S}$  & 7.24 & 5.25  & 7.01 & 4.53 &  6.01 \\
    $\text{Demucs}$ & $\text{W}$  & 6.84 & \textbf{7.01} & 6.86 & 4.42 &  6.28 \\
    $\text{ResUNetDecouple+}$ & $\text{S}$  & 8.98 & 6.04 & 6.62 & 5.29 &  6.73 \\
    $\textbf{MTFAttNet}$ (proposed) & $\text{S}$  & 
    \textbf{9.51} & 6.43 & \textbf{7.39} & \textbf{5.69} & \textbf{7.26} \\
    \midrule
    $\text{KUIELAB-MDX-Net}$ & $\text{W+S}$  &  \textbf{9.00} & 7.86 & 7.33 & \textbf{5.95} & 7.54 \\
    $\text{Hybrid Demucs}$ & $\text{W+S}$  &   8.04 & \textbf{8.67} & \textbf{8.58} & 5.59 & \textbf{7.72} \\
  \bottomrule
  \end{tabular}
  \vspace{0pt}
\end{table}
\begin{table}[htb]
  \caption{SDR comparison for different attention mechanisms.}
  \vspace{3pt}
  \label{SDR_Att}
  \centering
  \small
  \begin{tabular}{l c c c c c c}
    \toprule
   \textbf{Method} &\textbf{Vocals} & \textbf{Bass} & \textbf{Drums} & \textbf{Other} & \textbf{All} \\
    \midrule
    $\text{noAttNet}$ & 7.17 & 6.11 & 5.52 & 4.82 & 5.90\\
    $\text{FAttNet}$ & 8.42 & 6.19 & 6.44 & 5.56 & 6.65\\
    $\text{TAttNet}$ & 8.34 & 6.09 & 7.29 & 5.43 & 6.79\\
    $\text{TFAttNet}$ & 9.23 & 6.31 & 7.34 & 5.49 & 7.09\\
    $\text{MTFAttNet}$  & \textbf{9.51} & \textbf{6.43} & \textbf{7.39} & \textbf{5.69} & \textbf{7.26} \\
  \bottomrule
  \end{tabular}
\end{table}

\subsection{Attention mechanism study}
\label{ssec:AttStudy}
In this section, to better understand the benefit of the proposed MTFAttNet for MSS task, we further evaluate the systems by replacing the separator of MTFAttNet with different attention mechanisms in Table \ref{SDR_Att}. Condition TFAttNet employed the temporal-frequency attention with single scale attention structure illustrated in Fig.\ref{fig.Separator}(a). FAttNet and TAttNet are conditions only using the frequency attention module (with temporal attention removed) and the temporal attention module (with frequency attention removed), respectively. Condition noAttNet removed both temporal and frequency attention modules from condition TFAttNet.

We first discuss the overall performances of different attention systems. As shown in Table \ref{SDR_Att}, noAttNet achieves similar overall SDR (5.90 dB) to Spleeter (5.91 dB, listed in Table \ref{SDR_STOA}) indicating the effectiveness of proposed generic network structure without attention. With the frequency attention FAttNet and temporal attention TAttNet, the overall SDR is further improved to 6.65 dB and 6.79 dB respectively, which indicates the effectiveness of attention on capturing both temporal and frequency correlations. Combing both temporal and frequency attention, TFAttNet achieves the SDR of 7.09 dB. With multi-scale attention mechanism, MTFAttNet captures the spectrogram correlations more effectively and achieves the highest SDR of 7.26 dB.

The effects of attention mechanisms on different sources is varied for different type of stems. For the bass stems, we found that applying attention mechanisms introduced less improvements than other types of stems in Table \ref{SDR_Att}. This is potentially caused by the limited performance of the spectrogram domain method for the bass stems as shown in Table \ref{SDR_STOA}. Based on noAttNet, the additional attention modules can not effectively capture the spectrogram correlations for bass with limited frequency resolution.  
For drums, FAttNet achieves the SDR of 6.44 dB while TAttNet achieves the SDR of 7.29 dB. The temporal attention outperforms the frequency attention for drums, which might be due to the fact that drums contain repeated beats in temporal domain. 
For vocals, compared to condition noAttNet, both temporal and frequency attention can significantly improve the performance. Combing both temporal and frequency attention can increase SDR to 9.23 dB. With multi-scale attention, the SDR is further increased to 9.51 dB. The success of attentions on vocals might benefit from the relatively long duration of pitch and the harmonic structure of vocals, which leads to the high correlations in both temporal and frequency domain. The results show such temporal and frequency correlations can be effectively modeled by the proposed MTFAttNet system.

\section{CONCLUSION}

In this paper, we proposed a novel neural network architecture called MTFAttNet for music source separation. MTFAttNet employs a temporal-frequency attention module to exploit the spectrogram correlations along temporal and frequency dimension. A multi-scale mechanism is also proposed for the effectiveness of attention calculation. The experimental results show the proposed method achieves start-of-the art performance on the MUSDB18 dataset. In future work, we will explore the combination of waveform and spectrogram domain methods and further improve the separated results for instruments.

\bibliographystyle{IEEEbib}
\bibliography{icme2022template}

\end{document}